# Millimeter-Scale, Highly Ordered Single Crystalline Graphene Grown on Ru (0001) Surface

Y. Pan,[1] N. Jiang,[1] J.T. Sun,[1] D.X. Shi,[1] S.X. Du,[1] Feng Liu,[2] and H.-J. Gao[1*]

[1]*Institute of Physics, Chinese Academy of Sciences, Beijing 100080, China*
[2]*Depeartment of Materials Science & Engineering, University of Utah, USA*

**Abstract**

We demonstrate a method for synthesizing large scale single layer graphene by thermal annealing of ruthenium single crystal containing carbon. Low energy electron diffraction indicates the graphene grows to as large as millimeter dimensions with good long-range order, and scanning tunneling microscope shows perfect crystallinity. Analysis of Moiré pattern augmented with first-principles calculations shows the graphene layer is incommensurate with the underlying Ru(0001) surface forming a N×N superlattice with an average lattice strain of ~ +0.81%. Our findings offer an effective method for producing high quality single crystalline graphene for fundamental research and large-scale graphene wafer for device fabrication and integration.

Correspondence email: hjgao@aphy.iphy.ac.cn

The recent success in synthesizing single layer graphene [1,2] has fostered an exciting research field of graphene based materials stimulating many new discoveries [3-5] and potential applications [6-10]. Currently, there are three major methods for synthesizing single and few layer graphene: micromechanical cleavage or chemical exfoliation of highly oriented pyrolytic graphite (HOPG) [1,2,11,12], thermal decomposition of SiC [13,14], and epitaxial growth by vapour phase deposition of hydrocarbons or CO on metal substrates [15-18]. The graphene prepared by cleavage and exfoliation of HOPG has a very high quality of crystallinity as predefined in HOPG, but their dimensions are usually limited to μm dimension [1,2], and also with rather small productivity. Graphene layers formed on SiC surface usually contain multiple domains, poor long-range order, and structural defects [15]. Vapor phase deposition growth on metal substrates most often leads to graphene formation only on portion of the substrate surface [19]. Therefore, all the existing methods are limited in producing either high quality single crystalline graphene but with small dimension or large dimension but with poor long-range order and crystallinity. The limitation in sample size or quality hinders the measurement of certain physical properties, and will be the bottleneck to build graphene based electronic devices. There remains a strong need in developing graphene synthesis method to overcome these limitations. Here, we demonstrate successful growth of highly ordered single layer graphene on Ru(0001) surface up to an unprecedented large dimension of a few millimeters and with excellent crystallinity.

We grow graphene layers on Ru (0001) surface by thermal annealing. The experiment was carried out in a UHV chamber with the base pressure lower than $1\times10^{-10}$ mbar. The chamber was equipped with room temperature scanning tunneling microscope (RT-STM), Auger electron spectroscopy (AES), low energy electron diffraction (LEED), and electron beam heating (EBH) stage. The Ru crystal was a commercial product whose (0001) surface had been polished to less than 0.03 μm of roughness. It was cleaned by 3 cycles of ultrasonic cleaning in high purity acetone and ethanol to remove the organic contamination on the surface. Then it was loaded into the UHV chamber and pretreated using cycles of 0.8 keV Ar+ sputtering followed by annealing to 1350 K until sharp (1×1) LEED pattern and clean AES peaks of Ru without notable impurity peaks shown. After that the crystal was annealed in the following steps: slowly raising the temperature with pressure no higher than $1\times10^{-9}$ mbar, staying at 1000K for 20 min and slowly cooling down to room temperature. Afterwards, graphene had grown on the crystal surface, which was characterized *in situ* by LEED, STM, and AES, and further analyzed *ex situ* using low-temperature scanning tunneling spectroscopy (LT-STS) and x-ray photoelectron spectroscopy (XPS). The temperature was measured by a tungsten-rhenium (W-5%Re/W-26%Re) thermocouple welded to the EBH stage. LEED images were taken at the electron beam energy of 60 eV. The AES spectra were obtained with the electron beam energy of 3kV. The XPS and LT-STM analysis was done in a separate chamber. The crystal with graphene layer on its surface was taken out of the UHV chamber, protected by high purity Nitrogen in a bag while being delivered and loaded into the chambers with XPS facility and LT-STM, respectively.

Through the cycles of pretreatment in the UHV chamber, we have prepared highly ordered clean Ru (0001) surface, as confirmed by the sharp hexagonal (1×1) LEED pattern. After annealing, additional diffraction spots begin to appear indicating the formation of graphene layer on the Ru surface. In order to image the whole sample surface, we moved the sample and recorded the LEED pattern continuously, as shown in a video of the LEED screen [20]. Also, individual LEED patterns at different locations were taken at 1 mm interval across the surface. Figure 1 shows five examples of the LEED patterns. The two patterns taken at the sample edge (Fig. 1B and 1F) show almost no additional diffraction spots, indicating no graphene is formed in these edge areas. The three patterns taken in the inner region of the sample surface (Figs. 1C-1E) all show additional spots in the same pattern of arrangement, indicating graphene has formed in all these locations. Specifically, the new diffraction spots are arranged in a hexagonal pattern and in the same orientation as the Ru (1×1) spots. This kind of LEED pattern is the typical Moiré pattern resulted from the superimposition of two structures with different lattice constants and/or orientations [21]. From the measurement of the distances between the spots in LEED pattern, we deduce the ratio of lattice constant of graphene and Ru(0001) surface is about 12 to 11. Here, the most important finding is graphene has formed over a very large area of several millimeters on the Ru surface, as indicated by the sharp LEED patterns shown in Figs. 1C-1E.

To further reveal the detailed information of local surface structure, and assess the quality and continuity of the graphene flake, we have imaged the surface with the RT-STM, as shown in Fig. 2. Figure 2A shows the terraces were fully covered by atomic flat graphene. Each terrace is completely covered by a single domain of graphene and the domains on different terraces are orientated in the same direction and continuous. Figure 2B shows a higher resolution image, illustrating clearly the hexagonal Moiré pattern in real space. The average distance between the neighboring Moiré spots is 3 nm, which corresponds to twelve times of graphene lattice constant and eleven times of Ru lattice constant. Figure 2C is an even higher resolution image showing one unit cell of the Moiré pattern. The graphene overlayer was seen as a hexagonal lattice instead of a honeycomb lattice, because the image has a diatomic resolution [21]. In this image, twelve graphene lattices are counted in between the center of bright Moiré spots, i.e., the lattice parameter of the superlattice, consistent with the distance measurement in Fig. 2B.

The lattice mismatch between graphene and Ru (0001) surface drives the graphene overlayer into a corrugated surface forming a strained superlattice with an average in-plane tension of ~ +0.81%. Within each unit cell, the superlattice consists of three structural regions: the bright region (marked by the circle in Figure 2C) is bowed up into a ridge, the dark region (marked by dashed-line triangle) is bowed down into a valley, and the intermediate region (marked by dash-dotted-line triangle) has a medium height. In the valley region, the six-member C ring sits right on top of the six-member Ru ring underneath; while in the medium-height region, the six-member C ring sits atop above one Ru atom underneath. We have also measured STS at different locations in the surface. They all show a typical semi-metal behavior of graphene, as shown in the inset of Fig. 2A, which suggests a relatively weak interaction between the graphene overlayer and the underlying Ru surface.

One important finding is that the graphene can even form continuously over surface steps without breaking structure or symmetry, as shown in Fig. 3, suggesting the possibility of growing large scale graphene up to millimeter dimensions. We have taken the high-resolution STM images over the step edge areas. Figure 3A shows the Moiré pattern extending over several terraces and maintaining one single domain of superlattice passing over steps. Figure 3B is a close-up image of the step edge, which clearly shows that the graphene flake on the up terrace extends continuously to the lower terrace. Figure 3C shows at atomic-resolution STM image of the step edge, which further demonstrates that the atomic structure of graphene overlayer remains perfect crystallinity over a step without breaking bonds or defects. The crystalline continuity of the graphene over surface steps indicates that even though the width of the substrate terraces are only a few micrometers, the graphene 2D crystal can form with a much bigger dimension over several terraces, up to millimeter scale. This is found at least at many single-atomic-height steps, but we do observe graphene overlayer breaks up at double-atomic-height or higher steps.

In order to learn the element composition of the new structure, we performed AES and XPS analysis at room temperature. Figure 4A shows the AES spectra taken before and after annealing, i.e., graphene formation. The two peak positions at 230 eV and 273 eV remain the same, but the intensity of the peak at 273 eV is increased after annealing (red curve). This increase is attributed to the appearance of KLL peak of C at 271 eV upon graphene formation, which overlaps with the MNN peak of Ru at 273 eV. The ratio of peak heights, $R = (Ru_{273}+C_{271})/Ru_{230}$ can be used as an indicator of C accumulation on Ru surface. It is measured to be 2.1 before graphene formation, consistent with the typical R value of 1.85 to 2.05 for clean Ru surface [22], and 3.5 after. The XPS spectrum confirmed the sample contains no other element except for Ru and little amount of C, as shown in Fig. 4B. The C 1s peak (284.7 eV) overlaps with the Ru 3d peak (284.2 eV), but we can resolve the C 1s signal by subtracting the standard Ru XPS spectrum (red curve) from the measured spectra (black curve) and the remainder spectra gives rise to the C 1s peak at 284.7 eV (green curve).

In support of the experimental studies, we have carried out first-principles calculations to determine the ground-state structure of the graphene overlayer on Ru (0001). Our calculations were performed using DFT method, in which the generalized gradient approximation (GGA) [23,24] for exchange-correlation potentials, the projector-augmented-wave pseudo-potential approach and plane wave basis set for wave function expansion implemented in the Vienna Ab-initio Simulation Package (VASP) code [25] are used. The energy cutoff for plane wave is 400eV. We used a slab supercell containing 3-layer Ru atoms and monolayer graphene with a vacuum gap more than 16Å. The graphene and top two metal layers are allowed to relax in structural optimization until the atomic forces relax below 0.02eV/Å. All the calculations were done at the experimental Ru lattice constant (a=2.7058 Å and c= 4.2816 Å).

Figure 5A shows the optimized atomic structure with one unit cell of the strained graphene superlattice. One notices that in the middle of one half of the cell, the six-member C ring sits right on top of the six-member Ru ring underneath, as marked within the yellow triangle. In the middle of the other half, the six-member C ring sits atop above one Ru atom underneath, as marked within the blue triangle. This is in excellent agreement with the STM image shown in Fig. 2C. Figure 5B shows enlarged view of about two unit cells, illustrating the corrugated non-planar graphene surface, with surface heights coded in colors. The calculated height difference is about 0.4Å with the highest region coded in white color at the corner of cell in Fig. 5B. Figure 5C shows a larger view of the superlattice, which can be compared to the STM image in Fig. 2B. The calculations show that the lowest-lying C atoms are about 3.9Å above the underlying Ru surface, so likely the interaction between the graphene overlayer and the Ru surface is relatively weak, consistent with the suggestion by STS data.

The growth of graphene overlayer on the Ru surface is thermodynamically driven by a very large miscibility gap between Ru and C under 2000 $^o$C [26]. We confirmed that the carbon must come from the Ru bulk by segregating and accumulating on the surface during the annealing process, because any other possible carbon source was carefully excluded from the UHV chamber. However, the dimension, long-range order, and crystallinity of the graphene depend strongly on the kinetic parameters. If Ru was annealed at wrong temperature or for inappropriate time, the 2D hexagonal lattice is less ordered and contains high density of defects and dislocations. Large-scale high quality graphene can only be grown under the right growth conditions. The appropriate annealing temperature is 1000K. Initially when LEED pattern showed the graphene had formed, there were also huge clusters of several nm high on the surface, seen in the STM image as big protrusions. These amorphous C clusters form everywhere covering from 10% to 100% of graphene. But they can be removed by flash annealing the crystal in oxygen ambient at 800-1000K for 50-100s with an oxygen partial pressure of $5\times10^{-7}$ mbar.

In summary, we presented a large scale single layer graphene synthesized by thermal annealing of ruthenium single crystal. The grown graphene shows perfect crystallinity with good long-range order to as large as millimeter dimensions without breaking bonds over a step. Moiré pattern together with first-principles calculations shows the graphene layer is incommensurate with the underlying Ru(0001) surface forming a N×N superlattice with an average lattice strain of ~ +0.81%. Our findings offer an effective method for producing high quality single crystalline graphene for fundamental research on nanoelectronic devices, catalysis and Lithium ion based devices, as well as large-scale graphene wafer for device fabrication and integration.

We thank Prof. Qian Niu of the University at Austin for helpful discussion and suggestions, and H.M. Guo, Q. Liu, H.G. Zhang for experimental assistance. Work at IOP was supported by grants from National Science Foundation of China, National "863" and "973" projects of China, the Chinese Academy of Sciences, and Supercomputing Center, CNIC, CAS. Liu is supported by the DOE.

# FIGURE CAPTIONS

**Figure 1** LEED pattern at different locations of the sample. (A) The sample is a disk with a diameter of 8 mm, as shown in the photograph. The yellow line is the path of the e-beam when we move the sample to take the video of the LEED pattern. The red spots are the locations where we take the photos, as shown in (B)-(F). In (B) and (F), the additional spots are weak, but in (C), (D) and (E), the additional spots are sharp and in the same pattern of arrangement, indicating graphene has formed in all locations except the edge area.

**Figure 2** STM topograph of the graphene overlayer on Ru (0001) surface. (A) The atomic flat graphene flake extended over whole Ru terraces. (B) The hexagonal Moiré pattern formed by the superimposition of graphene and Ru substrate. (C) Atomic resolution image of one unit cell of the Moiré pattern. Tunneling parameters: (A) Vs=1.2V, I=0.17 nA (B) Vs=-1.2V, I=0.35 nA (C) Vs=-0.46V, I=0.27 nA.

**Figure 3** STM topograph of the graphene overlayer over steps. (A). The terraces are continuously covered by graphene as indicated by the Moiré pattern over steps. (B). A large view of the area enclosed by the green rectangle in (A), showing the continuous graphene flake on the step edges. (C) An atomic-resolution image shows the continuity of the atomic structure and crystallinity of the graphene overlayer across the substrate step edge. Tunneling parameters: (A) Vs=2.2V, I=0.13 nA (B) Vs=1.4V, I=0.17 nA (C) Vs=0.2V, I=0.43 nA.

**Figure 4** AES and XPS spectra. (A) AES spectra of Ru (0001) surface before (black curve) and after annealing (red curve). (B) XPS spectrum of Ru (0001) surface with graphene overlayer (black curve), which can be decomposed into the standard Ru spectrum (red curve) and the low intensity C spectrum (green curve).

**Figure 5** First-principles calculated graphene structure on Ru (0001). (A) Schematic illustration of the optimized atomic structure with one unit cell of the superlattice. (B) A large view of two unit cells showing surface corrugation. (C) An even larger view showing a perspective view of the superlattice.

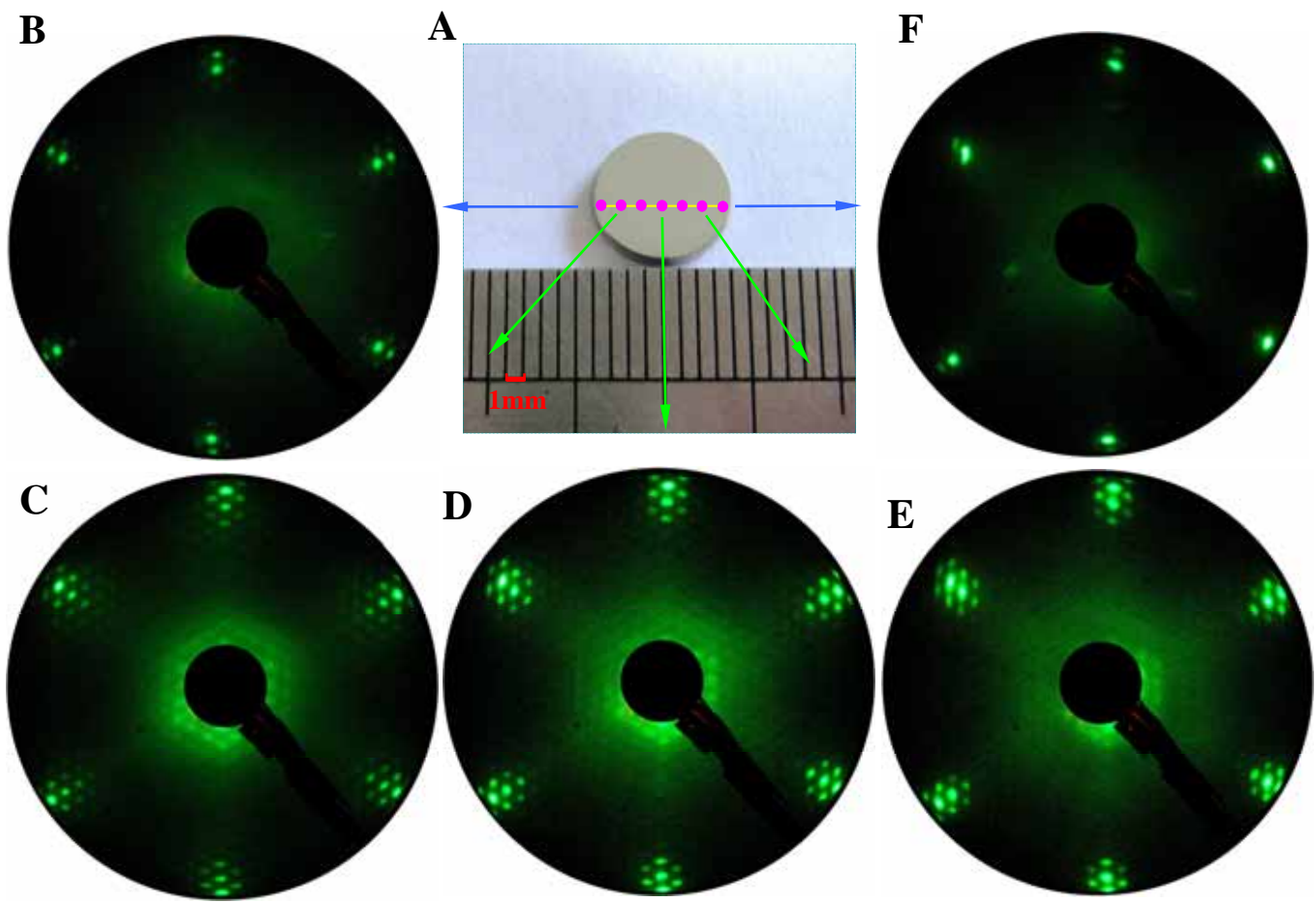

**Figure 1**

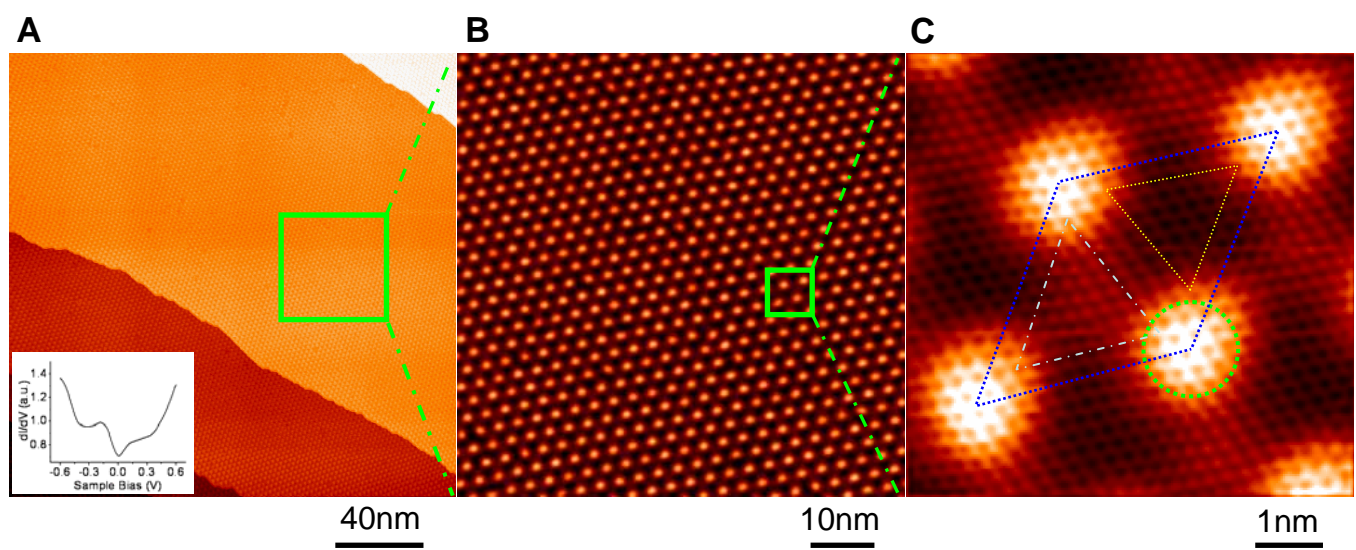

**Figure 2**

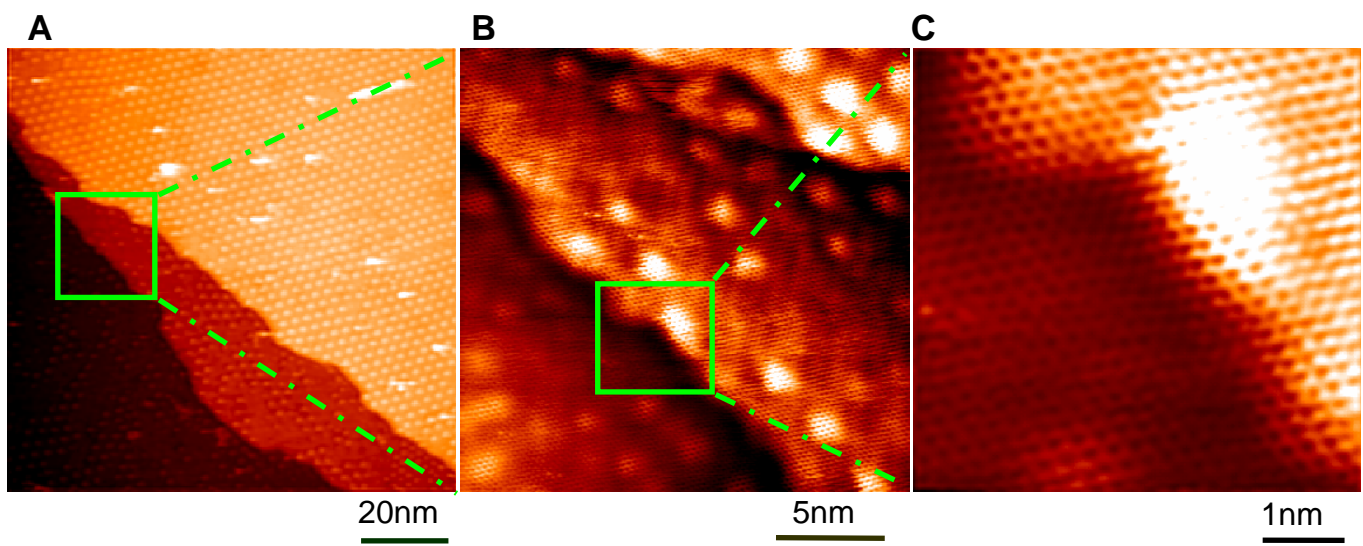

**Figure 3**

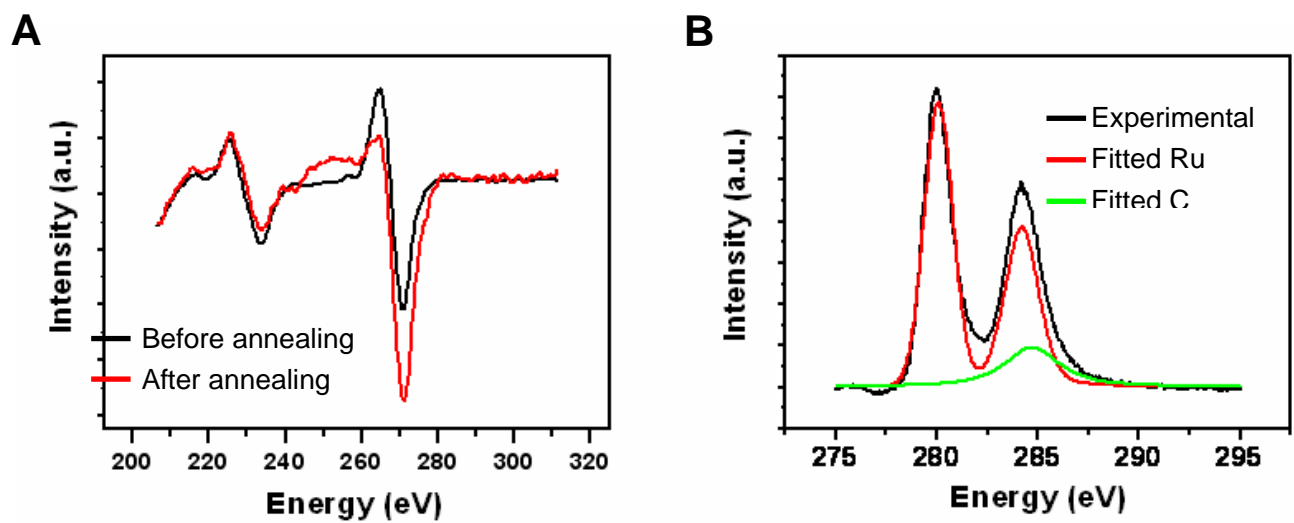

**Figure 4**

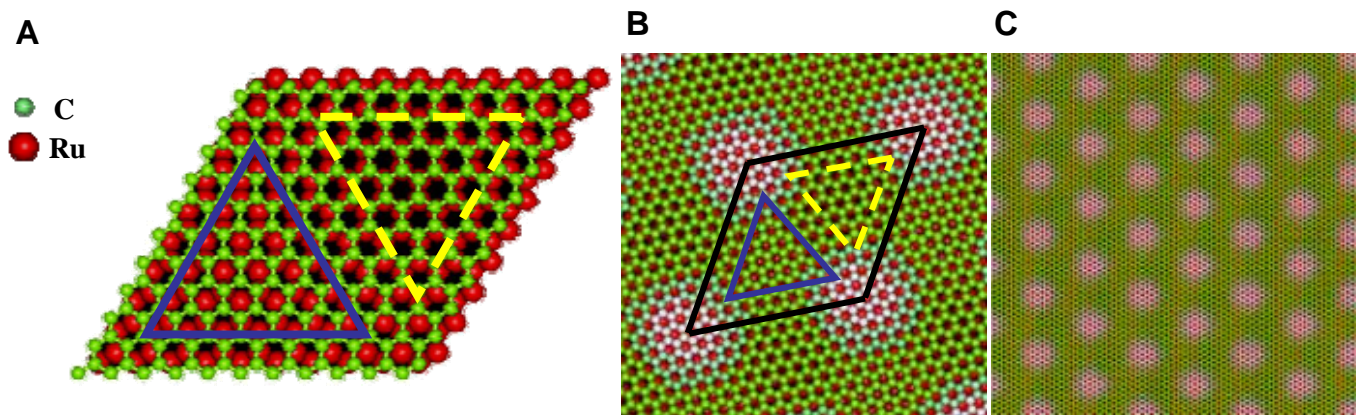

**Figure 5**